\documentclass[aip,rsi,preprint,nofootinbib]{revtex4-1} 
\usepackage{amssymb,amsmath,graphicx,centernot} 
\usepackage{dcolumn} 
\usepackage{bm} 
\usepackage{url} 
\usepackage{subcaption} 
\newcommand{\sgn}{\text{sgn}}
\begin{document}
\preprint{LA-UR-21-24726}
\title[Strength in Biaxial Compression]{Apparatus for Measuring Strength in Biaxial Compression}
\thanks{Email: katz@wuphys.wustl.edu}
\author{J. L. Belof}
\affiliation{Lawrence Livermore National Laboratory, Livermore, Cal.~94550}
\author{J. I. Katz}
\affiliation{Los Alamos National Laboratory, Los Alamos, N.~Mex.~87545}
\altaffiliation[Also at ]{Department of Physics and McDonnell Center for the Space Sciences\\Washington University, St.~Louis, Mo.~63130}
\date{\today}
\begin{abstract}
Most measurements of compressive strength of ductile materials have involved
Hopkinson-Kolsky bars or Taylor anvils placing samples in uniaxial 
compression.  In these geometries strain is limited by the tendency of the
sample to petal, in analogy to necking in uniaxial tension.  Estimation of
strength for any other form of the stress tensor requires assuming a shape
for the yield surface; because data exist only for uniaxial compression
these assumptions are untested.  In an imploding spherical shell compression
is biaxial, the plastic strain may not be small and the material behavior
may be nonlinear as a result of work hardening and heating by plastic work.
We propose to measure the strengths of materials in biaxial compression,
both quasistatically and dynamically, using the compression of thin
spherical shells.  We suggest surrounding the shell with an annulus filled
with a mixture of H$_2$ and Cl$_2$ gases whose volumetric detonation is 
initiated by a flash of blue and near-ultraviolet light.  Less promising
approaches are described in appendices.
\end{abstract}
\maketitle
\section{Introduction}
The strengths of materials in compression, particularly at higher strain
rates, are conventionally measured by the Hopkinson-Kolsky method
\cite{H14,K49,D48}.  A thin sample is placed between the flat ends of two
bars (an incident and a transmission bar) of strong material, and a pressure
pulse is transmitted through the incident bar to the test sample.  If the
pulse is strong enough, the sample, subjected to a uniaxial compressive
stress, flows plastically.  Its compressive yield strength is the minimum
compressive stress that produces plastic deformation.  Simple analysis
requires that the deformation be small, but small deformation is sufficient
to determine the stress threshold for plastic flow (the yield stress).
Alternatively, a flat-ended cylindrical bar of the test material impacts a
thick slab of much stronger material, called a Taylor anvil \cite{T48}, and
the threshold speed of impact at which the test material deforms plastically
determines its compressive strength.

To apply these results to stress states other than uniaxial compression
requires assuming a shape for the yield surface.  Many shapes have been
proposed; the most popular is probably that of von Mises:
\begin{equation}
\label{vonMises}
(\sigma_1 - \sigma_2)^2 + (\sigma_2 - \sigma_3)^2 + (\sigma_3 - \sigma_1)^2
= 2 S_u^2,
\end{equation}
where the $\sigma_i$ are the principal components of the stress tensor and
$S_u$ is the uniaxial yield strength (of a ductile material).  The validity
of this expression implies (or assumes) that the tensile and compressive
yield strengths are equal.  This is generally true for ductile materials,
but is wrong for brittle materials which have low tensile yield strengths as
a result of stress concentration at Griffith cracks; the von Mises
criterion assumes ductility and a homogeneous (on scales smaller than sample
feature sizes) stress field.  In pure shear the yield strength is
${1 \over 2}S_u$; this is obtained from the definition of shear stress
$\tau = (\sigma_1 - \sigma_3)/2$ where $\sigma_1$ and $\sigma_3$ are
respectively the major and minor principal stresses.

The validity of the von Mises yield surface is a fundamental question in
materials science.  Its simplest and most important test is to compare the
yield stresses in unconfined uniaxial and in biaxial compression.  The
corresponding geometries are the Hopkinson-Kolsky bar in uniaxial
compression and a thin spherical shell subjected to a uniform inward force
or impulse that create biaxial compression.  Once the yield strength is
exceeded the material flows plastically, and the relation between thei
applied forces and its final deformation is determined by its strength.
Measurement of its deformation and knowledge of the applied loads permit
determination of its strength.  We propose to apply this method to the
compression of thin spherical shells of ductile materials such as lead, tin
and copper.

A number of means of supplying momentum and energy to stress a spherical
shell and compress it may be considered.  These include explosives, sparks,
flyer plates, resistive heating of a driving fluid, gas pressure and
piezoelectrics.  Most of these methods suffer from difficulty in
maintaining the necessary symmetry; a thin test shell subject to asymmetric
loading will dimple, and even a symmetrically loaded shell may be subject to
buckling instability.

We have considered a number of methods of loading a spherical test object:
\begin{enumerate}
\item Piezoelectric drive is not strong enough to produce plastic flow in
all but the very weakest materials or in impractably thin shells.
\item Light initiated high explosives (LIHE), sprayed on as thin coats,
can be well matched to the strengths of transition metals, determining their
yield thresholds under dynamic biaxial compressive loads at interesting
values of strain rate.  However, it difficult to make these coats uniform to
better than 20\%, so that LIHE-driven thin shells are likely to crumple
rather than to retain their spherical geometry.
\item A liquid metal ``hydraulic fluid'' surrounding the test shell might be
pressurized by a spark or a tiny explosive or spark.  However, the resulting
weak shock would reverberate with very little damping as it propagates from
its point sources, and the test object would be loaded asymmetrically, and
be likely to crumple rather than to retain its spherical geometry.
\item We suggest that the most promising method is to surround the test
shell with an explosive mixture of H$_2$ and Cl$_2$, detonated homogeneously
by a flash of blue or near-ultraviolet light.
\end{enumerate}

Before describing these methods, we provide a general analysis of strength
measurements.  Then we describe the H$_2$-Cl$_2$ explosive method in detail.
The less promising methods 1--3 are described in appendices.
\section{Measurement of Uniaxial Strengths} 
Extant measurements of uniaxial compressive strengths are generally based
on data obtained from split Hopkinson bar, Taylor anvil (a hard surface
that stops a high velocity cylinder of the material tested), or shock
measurements.  In the first two methods an unconfined test object is 
subjected to uniaxial compression and flows plastically in the unconfined
directions.  The achievable strain is limited by the finite tensile strain
at rupture of even ductile materials.  In uniaxial compression at high
strains the plastically deforming material undergoes a petaling instability
in which azimuthal asymmetry grows until the outer parts of the test object
break into separate petals.  The strain diverges at the apex of an opening
crack between the petals but remains small within them.

In order to achieve a yield stress $\sigma$ the longitudinal velocity in a
Hopkinson bar or Taylor anvil experiment must be
\begin{equation}
v_l = {\sigma \over Z_l},
\end{equation}
where $Z_l = \rho c_l$ is the longitudinal acoustic impedance and $c_l$ the
unconfined longitudinal wave speed.  At higher $v_l$ the material flows
plastically; $\sigma$ may increase as a result of work hardening, or
decrease as a result of heating by plastic work.  The characteristic time
for a strain $\epsilon$ and displacement $\epsilon r$ of a cylinder of
radius $r$ is
\begin{equation}
t_{char} = {\epsilon r \over v_l} \sim {\epsilon r Z_l \over \sigma}.
\end{equation}
The corresponding strain rate is
\begin{equation}
{\dot \epsilon} \sim {\epsilon \over t_{char}} \sim {\sigma \over r Z_l}.
\end{equation}
Representative numbers for a transition metal are $\sigma = 3$ kbar,
$r = 0.1$ cm, $\rho = 8$ gm/cm$^3$ and $c_l \approx c_s \approx 3$ km/sec,
yielding a characteristic
\begin{equation}
{\dot \epsilon} \sim 10^4\ {\rm s}^{-1}.
\end{equation}
This is a practical upper limit on achievable values of $\dot \epsilon$.
For very soft metals like indium the corresponding limits on $\dot \epsilon$
may be smaller by as much as two orders of magnitude.

The strength of ductile metals is generally an increasing function of their
strain rate \cite{N81,F86,HC87,LLK98,MPB01}, particularly at strain rates
$>10^3$ s$^{-1}$.  High strain rates $\dot \epsilon$ imply high dislocation
velocities \cite{MPB01}
\begin{equation}
v_{disc} = {{\dot \epsilon} \over n_{disc} a},
\end{equation}
where $n_{disc}$ is the dislocation density and $a$ is an interatomic
spacing.  The dislocation density may be as small as $\sim 10^7$/cm$^2$ in
annealed metals or as large as $\gtrsim 10^{11}$/cm$^2$ in cold-worked metals
\cite{WS56}.  A low-melting soft metal like indium ($T_m = 430$ K) is likely
to anneal at room temperature and to have $n_{disc}$ in the lower end of
this range.  Even Sn ($T_m = 505$ K) and Pb ($T_m = 600$ K) recrystallize
at room temperature, although not on the ms time scale of a dynamic
experiment.  As $v_{disc}$ increases, phonon drag on the dislocation
increases rapidly and replaces pinning as the principal impediment to
dislocation motion.  This emphasizes the importance of obtaining strength
data at strain rates of interest because the dependence of strength on
strain rate is significant.

In shock experiments \cite{SCG80,SL89,R03} the pullback from shock
reflection at a free surface measures the tensile strength of the
material in its post-shocked (heated and damaged by plastic flow) state.
The strain rates are subject to similar limitations as those in
compressional experiments because the material velocities are similarly
limited if it is required that the shock not far exceed the spall limit.
The test rod thickness is replaced by the slab thickness.  Shock
experiments are, in general, harder to interpret quantitatively than 
unshocked experiments on bars.
\section{Effect of Geometry}
The stress field in a split Hopkinson bar experiment is uniaxial compression
(unconfined in the the two transverse dimensions), in a shock experiment it
may be uniaxial and slab-symmetric (transversely confined on the
experimental time scale) tension, and in a converging thin spherical shell
it is biaxial compression (unconfined in the radial direction).  As a
result, these are not directly comparable, and none may be directly
applicable in other circumstances.

The stress tensor at yield in uniaxial compression, assuming the von Mises
yield criterion (Eq.~\ref{vonMises})
is
\begin{equation}
\label{uniaxialstress}
\mbox{\boldmath$\sigma$} = S_u\ \sgn{(P)} \left\{
\begin{vmatrix}
1 & 0 & 0 \\
0 & 0 & 0 \\
0 & 0 & 0 
\end{vmatrix}
= 
{1 \over 3}\ \mbox{\boldmath$1$}
+ 
\begin{vmatrix} 
{2 \over 3} & 0 & 0 \\
0 & - {1 \over 3} & 0 \\
0 & 0 & - {1 \over 3} \\
\end{vmatrix}
\right\}
,
\end{equation}
where $P$ is the applied uniaxial pressure
\cite{W64}\footnote{{\it N.~B.:\/} $S_u$ may be an increasing function of
the isotropic pressure, the first Cauchy invariant $I_1$.  Such a dependence
in granular materials, in which friction resists plastic flow and that get
much stronger under compression, is described by a Mohr-Coulomb yield
surface.  In ductile materials, for $I_1 \ll K$ ($K$ is the bulk modulus)
the density is close to that of the uncompressed state and the dependence of
$S_u$ on $I_1$ is expected to be weak.}.

The stress tensor at yield of a thin spherical shell of thickness $\delta r$
and radius $r_0$ loaded by an external pressure $P$, to lowest order in 
$\delta r/ r_0$ and assuming the von Mises yield criterion, is
\begin{equation}
\mbox{\boldmath$\sigma$} \approx S_u\ \sgn{(P)} \left\{
\begin{vmatrix}
0 & 0 & 0 \\
0 & 1 & 0 \\
0 & 0 & 1
\end{vmatrix}
= 
{2 \over 3}\ \mbox{\boldmath$1$}
+ 
\begin{vmatrix} 
-{2 \over 3} & 0 & 0 \\
0 & {1 \over 3} & 0 \\
0 & 0 & {1 \over 3} \\
\end{vmatrix}
\right\}
.
\label{shellstress}
\end{equation}
The components $\sigma_{\theta\theta} = \sigma_{\phi\phi} \approx P r_0 /
(2 \delta r)$, while $\sigma_{rr} = {\cal O}(P)$ \cite{LL59}.  According to
the von Mises criterion, yield begins at 
\begin{equation}
P \approx {2 \delta r \over r_0} S_b \ll S_b,
\end{equation}
defining the biaxial yield stress $S_b$.  Eq.~\ref{vonMises} implies that
$S_b = S_u$ and that these yield strengths are the same in tension as in
compression.  The compressive stress is biaxial to lowest order in $\delta r
/ r_0$.

The deviatoric parts of the stress tensor differ in these two geometries.
For compressive loads ($P > 0$) two components are negative in uniaxial
compression, while only one is negative in biaxial compression.  The
compressive isotropic part of the stress tensor is twice as large in
spherical compression, and acts to prevent petaling or spall.  In the
plastic flow of a bar in longitudinal compression the circumferential and
radial length elements grow, while in the spherical compressive flow of a
shell all non-radial length elements shrink and only the one radial
dimension of length element grows.

A ductile compressively loaded bar petals at large strains, similarly to the
necking of a bar in longitudinal tension.  Stress and flow are greatest
where the material is thinnest, leading to further thinning, increasing
strain where it is thinnest, and finally to rupture if the limiting plastic
strain is reached.  In comparison, in the spherical shell only radial length
elements stretch, and the opening of cracks and voids is prevented both by
the (small) radial compressive load and the greater isotropic compressive
stress.


The von Mises yield surface (Eq.~\ref{vonMises}) is an assumption that
requires empirical justification.  Is the value of yield stress $S_u$ found
for uniaxial compression also applicable to biaxial compression, as implied
by Eq.~\ref{vonMises}?  Two dimensional compression of a thin spherical
shell is expected to be immune to petaling instability, and in the plastic
flow regime immune to elastic buckling instability.  By comparison to the
strength measured in one dimensional compression it can provide a direct
test of the validity of the von Mises yield surface.  It can also measure
such phenomena as the strain rate dependence of the strength and work
hardening.

\section{Spherical Shells}
\label{thinshell}
We consider the dynamics of a spherical shell impulsively launched inward
whose initial kinetic energy is dissipated as heat by plastic work.  In
general, this is a nonlinear problem that requires numerical calculation.
Here we present analytic results for the simplest possible case, a shell
that remains thin, without its strength changing as a result of work
hardening or plastic heating. 
\subsection{Motion}
First consider a thin spherical shell of incompressible matter of mass $M$,
density $\rho$, mean radius $r$ and thickness $\delta r \ll r$, undergoing
radial flow with slowly varying kinetic energy $E$.  Its mean mass-averaged
radial velocity $v_m$ also varies slowly.  However, its thickness changes
with time.  To lowest order in $\delta r/r$:
\begin{equation}
\delta r \approx {M \over 4 \pi \rho r^2},
\end{equation}
and
\begin{equation}
v_m \approx \sqrt{2 E \over M}.
\end{equation}

The speeds of its surfaces are
\begin{equation}
v_\pm \approx v_m \left(1 \mp {\delta r \over r}\right),
\end{equation}
where the upper signs refer to the outer surface and the lower signs to
the inner surface.

The accelerations of its surfaces are
\begin{equation}
a_\mp \approx \pm 3\ \sgn(v_m){\delta r \over r} {v_m^2 \over r}.
\end{equation}
In an imploding flow the speed of the inner surface increases and that of
the outer surface decreases.  In an expanding flow the opposite is true.
\subsection{Plastic work}
\label{plasticwork}
For a thin spherical shell that has undergone a change in radius $\delta r$
in radial incompressible flow, the diagonalized strain tensor in local
cartesian coordinates of an element
\begin{equation}
\label{shellstrain}
{\mathbf u} = 
\begin{vmatrix}
{d \delta r \over \delta r} & 0 & 0 \\
0 & - {d \delta r \over 2 \delta r} & 0 \\
0 & 0 & - {d \delta r \over 2 \delta r}
\end{vmatrix}
.
\end{equation}

Using the expressions for stress (\ref{shellstress}) and strain 
(\ref{shellstrain}), the plastic work (ignoring the elastic deformation) per
unit volume of the shell
\begin{equation}
dW = {\mathbf u} \cdot \mbox{\boldmath$\sigma$} = - S_b {d \delta r \over
\delta r}.
\end{equation}
Integrating, noting that $\delta r \propto R^{-2}$ where $R$ is the shell's
radius,
\begin{equation}
\label{worksphere}
W = \int\! dW = 2 S_b\ \vert\ln{(R_i / R_f)}\vert,
\end{equation}
where $R_i$ and $R_f$ are the initial and final radii.

\section{Biaxial Strength}
Yield strength in biaxial compression is of intrinsic interest because it
offers the opportunity to test models of the yield surface.  This geometry
can also enable measurement of strength at high strain (because of the
absence of the petaling instability that occurs in uniaxial compression) and
at high strain rate.
\subsection{Quasistatic Compression}
\begin{figure}
\begin{center}
\includegraphics[width=5in]{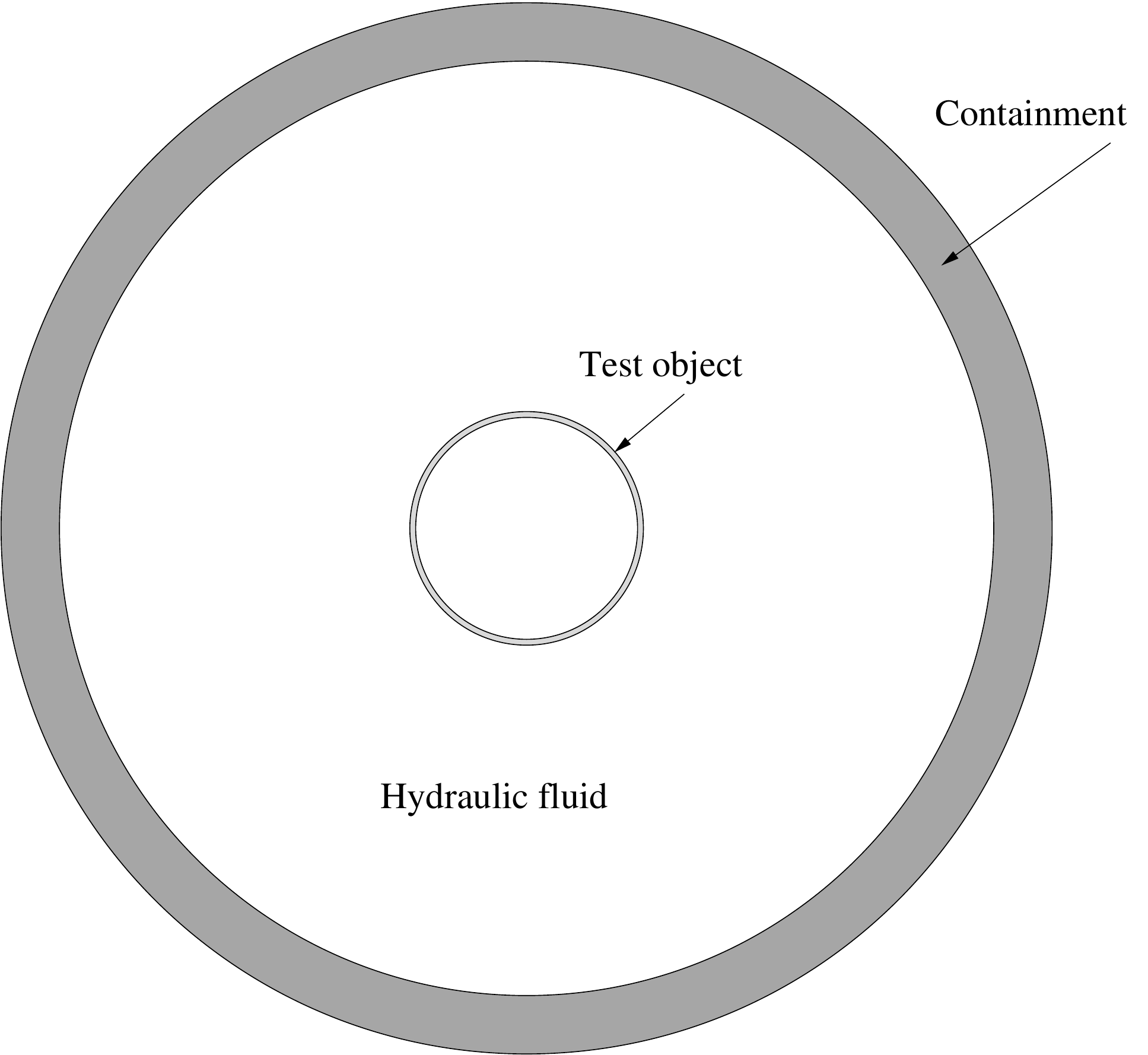}
\caption{Schematic of quasi-static biaxial compression experiment.}
\label{exptfig1}
\end{center}
\end{figure}

The strength of a thin shell in biaxial compression at low strain rates can
be measured using hydraulic compression.  A thin spherical shell of the
material to be tested is placed in a chamber filled with hydraulic fluid.
This chamber is thick-walled with a stiff material such as steel, so that
its deformation is negligible.  If the test object has a small volume
compared to that of the chamber then the volume of the chamber remains
essentially constant even as the test object shrinks to smaller volume (the
shrinkage is of its central void; the metal is essentially incompressible).
A schematic diagram of the experiment is shown in Fig.~\ref{exptfig1}.

The stress in the test object
\begin{equation}
\label{shelleq}
\sigma = {p R \over 2 \delta r},
\end{equation}
where $p$ is the external pressure, $R$ is its radius and $\delta r \ll R$ is
the thickness of its shell.  As it compresses at constant $p$ and constant
density, the shell thickens 
$\delta r \propto R^{-2}$, so the pressure threshold for plastic flow
\begin{equation}
\label{pthresh}
P_{thresh} = {2 S_b \delta r \over R} \propto R^{-3},
\end{equation}
where $S_b$ is the biaxial compressive yield stress of the shell.  The
increase of $P_{thresh}$ with decreasing $R$ makes compression stable,
provided the shell remains spherical.  Compression is quasi-static as the
pressure of the hydraulic fluid is increased\footnote{Expanding the shell
with an internal gas is only neutrally stable on the (slow) time scales on
which the gas is isothermal because $P_{thresh} \propto V^{-1}$ while the
internal pressure of an isothermal gas is also $p \propto V^{-1}$.  On
faster (dynamic) time scales $p \propto V^{-\gamma}$, where $\gamma > 1$ is
the adiabatic index, and is stable provided spherical symmetry is
maintained.  In fact, expansion by an internal fluid is unstable to
dynamical non-spherical instability: any weaker or thinner part of the shell
will stretch preferentially, with the expansion occurring disproportionately
there.  Catastrophic rupture will be almost immediate.  It is only possible
to inflate a rubber balloon without rupture because of the nonlinear
stress-strain relation of rubber: After a 2--3 fold stretch it becomes much
stiffer.}.  Work-hardening may make $P_{thresh}$ increase even faster than
$\propto R^{-3}$.

In order to maintain, to a good approximation, a constant pressure during
this process, the hydraulic pump (not shown in the Figure) must respond to
the shrinkage of the test object on the time scale on which it shrinks.
Because an infinitesimal overpressure produces an infinitesimal shrinkage,
provided the plastic flow of the test object is ductile, this condition is
expected to be met.  

It may be attractive to use very thin shells to minimize the pressure
required to produce plastic work (Eq.~\ref{pthresh}).  However, if the shell
is too thin the possibility of buckling must be considered.  Buckling of a
elastic thin perfectly spherical shell occurs at a pressure \cite{LL59}
\begin{equation}
\label{buckle}
P_{buckle} = {2 E \over \sqrt{3(1-\nu^2)}}\left({\delta r \over R}\right)^2,
\end{equation}
where $E$ is the Young's modulus and $\nu \approx 0.3$ the Poisson's ratio
of the shell.  Combining this with Eq.~\ref{pthresh} yields the condition
that the shell not buckle at its plastic yield threshold
\begin{equation}
\label{thinness}
{\delta r \over R} > {S_b \over E} \sqrt{3(1-\nu^2)} \sim
0.001\text{--}0.005,
\end{equation}
where typical parameters for transition metals have been used.  However,
such very thin shells may buckle as a result of deviations from perfect
sphericity and may be difficult to handle without damage.  For these
practical reasons it will probably be necessary to use shells at least ten
times thicker than those permitted by Eq.~\ref{thinness}; this equation then
demonstrates that spherical elastic buckling will not occur.
\subsection{Dynamic Compression}
\label{dynamic}
Consider an impulse per unit area $p_f$ applied to a thin spherical shell of
the sample material of density $\rho$, initial radius $R_i$ and thickness
$\delta r$.
Equating the kinetic energy per unit volume ${p_f^2/2\rho(\delta r_0)^2}$,
where $\delta r_0$ is the initial shell thickness, to the plastic work
(\ref{worksphere}) done before the shell stops at a final radius $R_f$,
yields a relation between its yield strength $S_b$ and $R_i/R_f$, with the
other variables as parameters:
\begin{equation}
\label{Y0}
S_b = {p_f^2 \over 4 \rho (\delta r_0)^2 |\ln{(R_i/R_f)}|}.
\end{equation}
It is probably best to choose an impulse so that $\ln{(R_i/R_f)} \lesssim
0.5$, preserving the spherical geometry of the shell even if the implosion
is not very accurately spherical.  For $S_b = 1$ kbar (a plausible value for
a transition metal), $\delta r_0 = 1$ mm, $R_f = 0.6 R_i$ and $\rho = 8.9$
g/cm$^3$ (copper) and defining $\delta \equiv (R_i - R_f)/R_i$, if $\delta
\ll 1$ the required impulse per unit area
\begin{equation}
\label{pf}
p_f \approx 1.9 \times 10^4 \sqrt{\delta}\ \text{taps},
\end{equation}
where 1 tap $\equiv$ 1 g cm$^{-1}$ s$^{-1}$ is the cgs unit of impulse per
unit area.

The radial strain rate varies during the trajectory of the shell, but may be
approximated
\begin{equation}
{\dot \epsilon}_r \approx {\epsilon_r \over t_{char}} \approx {2 p_f \over
R_i \rho \delta r_0} \approx 4 \sqrt{S_b \delta \over \rho}{1 \over R_i}
\approx 2 \times 10^4\ \text{s}^{-1},
\end{equation}
where the radial strain $\epsilon_r \approx 2 (R_i-R_f)/R_i$, the
characteristic time $t_{char} \approx (R_i - R_f)/v$ and the initial
velocity $v = p_f/(\rho \delta r_0)$.  The final result uses Eq.~\ref{pf}
for the impulse per unit area $p_f$ and the numerical values $S_b = 1$ kbar,
$\rho = 8.9$ g/cm$^{3}$, $\delta = 0.3$ and $R_i = 1$ cm.  The strains and
strain rates in the two tangential dimensions are half as large.
\section{Experimental Method: Hydrogen-Chlorine Explosions}
\begin{figure}
\centering
\includegraphics[width=5in]{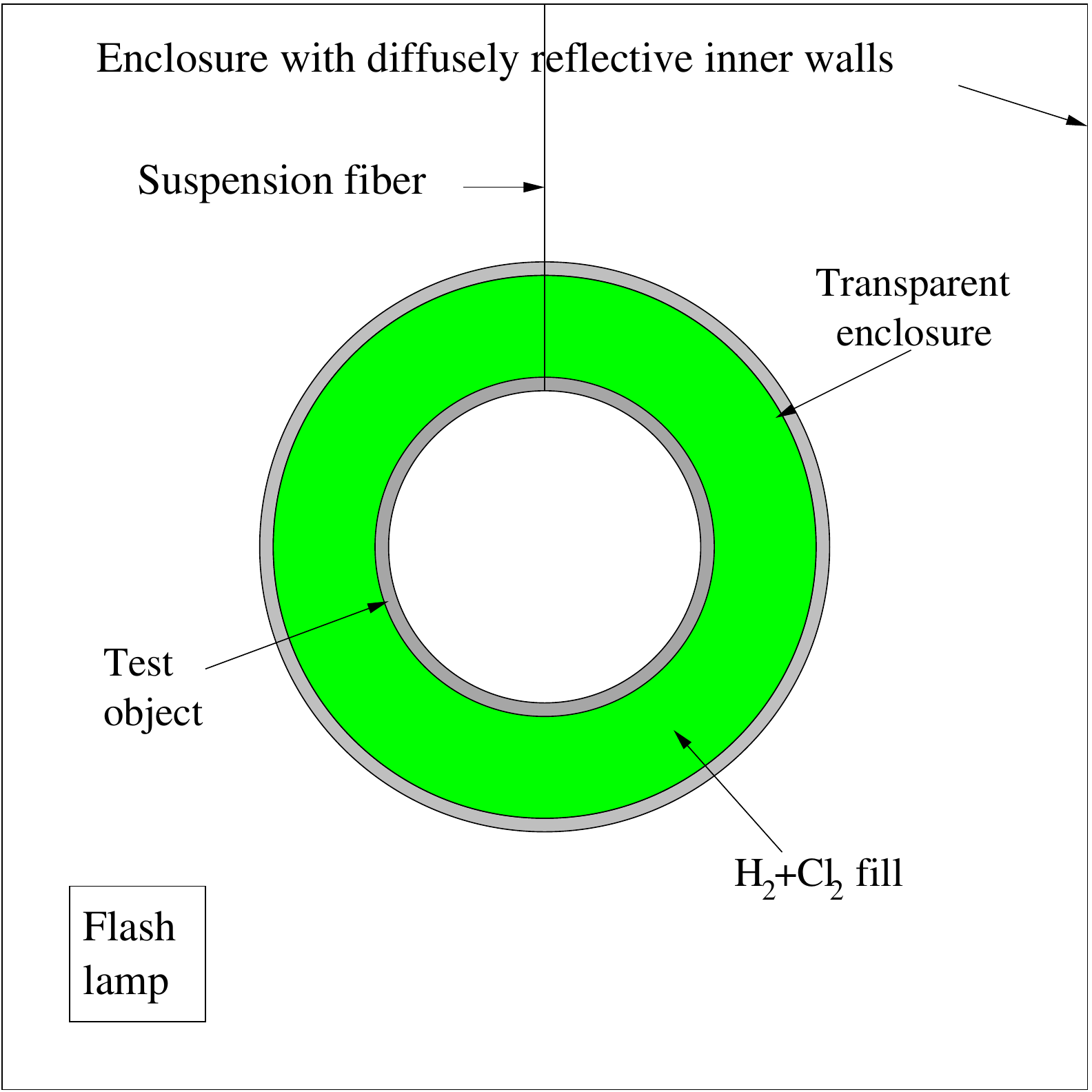}
\caption{\label{HCl} Photo-induced detonation of a mixture of H$_2$ and
Cl$_2$ applies a very rapidly rising pressure to the test object.  Flooding
the enclosure with blue or near-ultraviolet light initiates all the
explosive mixture simultaneously, ensuring symmetry of the applied pressure,
provided the transparent containment vessel is spherical and concentric with
the test object.  If the containment vessel is frangible, the suddenly
released pressure will fragment it into many small fragments, and the
high pressure HCl gas will rapidly escape between the fragments, reducing
the pressure by a large factor.  The load on the test object will be nearly
impulsive on the time scale of its motion.}
\end{figure}

The explosive reaction of hydrogen and chlorine 
\begin{equation}
\label{HClreact}
\mathrm H_2 + Cl_2 \to 2 HCl
\end{equation}
may be initiated by blue or near-ultraviolet light, and is explosive.  This
suggests a means of applying a nearly impulsive pressure to a test shell, as
shown in Fig.~\ref{HCl}.
The reason for choosing a hydrogen-chlorine reaction rather than the more
familiar explosive hydrogen-oxygen reaction
\begin{equation}
\label{HOreact}
\mathrm 2 H_2 + O_2 \to 2 H_2 O
\end{equation}
is the ease of photoinitiating the hydrogen-chlorine explosion, permitting
the simultaneous and uniform development of overpressure.  The
hydrogen-oxygen explosion is generally initiated by a spark, and a
detonation front propagates in a complex and asymmetric matter from the
point of initiation.

Blue or ultraviolet light dissociates Cl$_2$ molecules, leading to the
chain reaction
\begin{eqnarray}
\mathrm Cl + H_2 &\to HCl + H \label{step1}\\
\mathrm H + Cl_2 &\to HCl + Cl \label{step2}.
\end{eqnarray}
Reaction \ref{step1} is endothermic by an energy (in temperature units) of
about 600 K, small enough that it is readily supplied by thermal energy at
room temperature, while reaction \ref{step2} is exothermic by about 22740 K,
yielding a net enthalpy of reaction of about 11070 K per HCl molecule. 
These exchange reactions have large cross-sections $\sigma \sim 10^{-16}$
cm$^2$.

This is not a branching chain, and may be terminated by reactions
\begin{eqnarray}
\mathrm Cl + Cl + R &\to Cl_2 + R\\
\mathrm H + H + R &\to H_2 + R.
\end{eqnarray}
The bystander $R$ is necessary because the rate of radiative association
is very small in these reactions that have no dipole moment.  As a result,
it is necessary that a non-infinitesimal fraction of the Cl$_2$ be
photodissociated.  Once a significant amount of product HCl has been
produced, the temperature rises sufficiently that the chain branches by
the reaction
\begin{equation}
\mathrm Cl_2 + R \to Cl + Cl + R,
\end{equation}
which is endothermic by 29120 K (2.51 eV), and rapidly proceeds to
completion.

The characteristic reaction time at an initial pressure $P$
\begin{equation}
t_{char} = {1 \over n \sigma_{react} v_{th}} \sim 3 {1\,\text{bar} \over P}
\ \text{ns},
\end{equation}
where we have taken a reaction cross-section $\sigma \sim 10^{-16}$ cm$^2$,
$n$ is the reactant density and $v_{th} \sim 10^5$ cm/s is the thermal
velocity of the hydrogenic species at room temperature.  Once the chain is
initiated, it rapidly proceeds to completion.  These reactions have been
studied in detail because of their significance for the chlorine industry
in which electrolytically produced chlorine is mixed with some hydrogen;
see \cite{CI} for a bibliography.
\subsection{Radiative Initiation}
\begin{figure}
\centering
\begin{subfigure}[b]{0.49\textwidth}
\includegraphics[width=\textwidth]{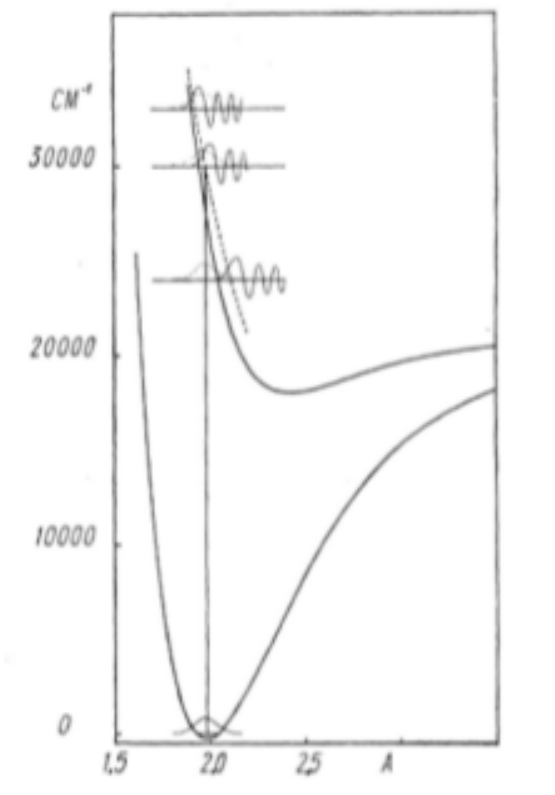}
\end{subfigure}
\begin{subfigure}[b]{0.49\textwidth}
\includegraphics[width=\textwidth]{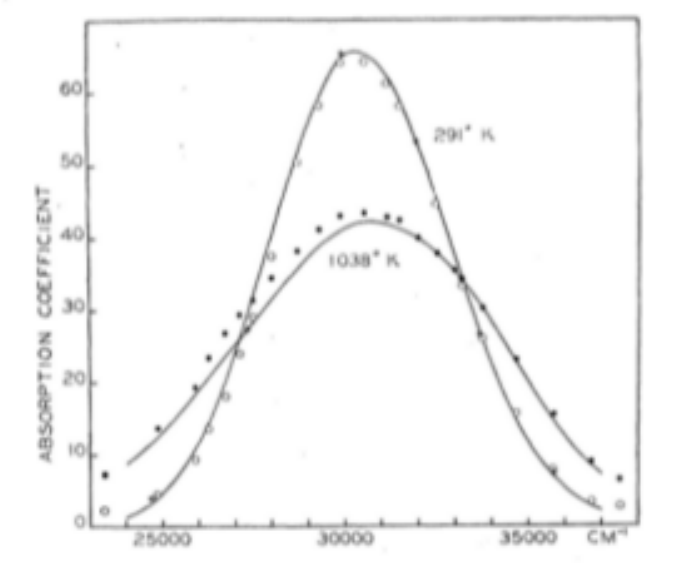}
\end{subfigure}
\caption{\label{HClphoto} Left subfigure shows the level diagram of the
Cl$_2$ molecule.  The solid lines show the Morse potentials of the ground
X $^1\Sigma_g^+$ state and the first excited $^1 \Pi_u$ state with an
allowed transition.  Because of the Franck-Condon principle, excitation from
the ground state has large matrix element only into unbound high vibrational
levels of the excited electronic state at shorter wavelengths than the
$\lambda = 4915\,$\AA\ corresponding to the dissociation threshold of 2.51
eV.  The right subfigure shows the absorption coefficient as a function of
reciprocal wavelength, in units of (dex/cm)/(moles/l) at 291 K; multiply by
0.1004 to transform to cm$^{-1}(P_{\text{Cl}_2}/\text{bar})$ for Cl$_2$
partial pressure $P_{\text{Cl}_2}$ at 298 K \cite{GB33,GRB33}.}
\end{figure}

In order to assure uniform volumetric initiation it is necessary that the
entire volume be illuminated.  The level diagram of the Cl$_2$ molecule and
its photoabsorption cross-section are shown in Fig.~\ref{HClphoto}.

Near the near-ultraviolet absorption peak, the attenuation at
$P_{\text{Cl}_2} = 5$ bars (as discussed in Sec~\ref{pressure} pressures in
this range are relevant) is $\sim 25$ cm$^{-1}$, so the illumination would
not penetrate very deep into the gas.  However, the attenuation at longer
(blue and green) wavelengths is much less, so this light from a
broad-spectrum source would penetrate through the gas and initiate the
reaction in the inner regions of the gas-filled annulus.  The product HCl is
transparent at visible and near-ultraviolet wavelengths with $\lambda > 2270$
\AA\ \cite{HCldata}.
\subsection{Pressure Multiplication}
\label{pressure}
In order to force a shell of thickness $\delta r$ and initial radius $R_i$
to flow plastically, in a static approximation a pressure $P = 2 Y \delta r/
R_i$ is required, where $Y$ is the yield stress.  For plausible parameters
for a soft transition metal $Y = 1$ kbar and $\delta r/R_i = 0.05$; $P =
100$ bars would be required.  The criterion is more complicated for
dynamic loading (Sec.~\ref{dynamic}).

There is a simple relation between the gas pressure before ignition of a
stoichiometric mixture and afterward:
\begin{equation}
{P_f \over P_i} = {\Delta H \over k_B T_i} {2 \over n},
\end{equation}
where $\Delta H$ is the enthalpy of reaction, $T_i$ the initial gas
temperature, $n$ the number of effective degrees of freedom of the burned
gas and the molar ratio is 1 for the reaction $\mathrm H_2 + Cl_2 \to 2HCl$
but 2/3 for the reaction $\mathrm 2H_2 + O_2 \to 2H_2O$.  The enthalpy for
the chlorine reaction is 92.34 kJ/mole (HCl), while for the oxygen reaction
it is 242 kJ/mole (H$_2$O); these values assume complete reaction, which is
in fact an overestimate by about 15\% for the chlorine reaction and by more
for the oxygen reaction.  For HCl $n \approx 6.5$ (Sec.~\ref{dof}), while
for H$_2$O, a nonlinear triatomic molecule produced with large $\Delta H$,
all nine degrees of freedom may be approximated as classically excited.  The
resulting pressure ratios (with $T_i = 298$ K) are
\begin{eqnarray}
\text{HCl} &\quad {P_f \over P_i} = 11.5 \sim 10 \label{pHCl}\\
\mathrm{H_2O} &\quad {P_f \over P_i} = 21.7 \sim 10, \label{pH2O}
\end{eqnarray}
where the final approximate values allow for the fact, discussed in
Sec.~\ref{completeness}, that the reactions don't go to completion because
of the high temperatures of the product gases.  These reactions, explosively
turning gases into other gases, may be thought of as pressure multipliers.
The greater $\Delta H$ of the oxygen reaction is partially offset by the
greater number of degrees of freedom of its products.
\subsubsection{Degrees of Freedom}
\label{dof}
At the temperatures produced by reaction (\ref{HClreact}), a few thousand K,
the quantized nature of the vibration of HCl, with a level spacing of 2991
cm$^{-1}$ \cite{HCldata} (implicitly multiplied by $hc$ so that 1 cm$^{-1}$
corresponds to $1.986 \times 10^{-16}$ erg), or 4304 K in temperaturei
units), is important.  The total vibrational energy per molecule is
\begin{equation}
{\cal E} = h \nu e^{-h\nu/k_B T} + 2 h \nu e^{-2h\nu/k_B T} + \ldots =
{e^{-h\nu/k_B T} \over \left(1 - e^{-h\nu/k_B T}\right)^2}.
\end{equation}
The energy balance equation
\begin{equation}
\Delta H = 5k_B T + 2{\cal E},
\end{equation}
where the first term is the contribution from the three translational and
two rotational degrees of freedom and $\Delta H$ describes reaction
(\ref{HClreact}) with two moles of HCl on the right hand side, may be solved
numerically (by successive approximation) for $T$.  Then the effective
number of vibrational degrees of freedom is
\begin{equation}
n_{vib} = 2 {{\cal E} \over k_B T}
\end{equation}
and
\begin{equation}
n = 5 + n_{vib}.
\end{equation}
\subsubsection{Completeness of Reaction}
\label{completeness}
For the reaction (\ref{HClreact}) the final temperature, ignoring the
(small) contribution of the initial thermal energy, is simply related to
$\Delta H$:
\begin{equation}
k_B T = {\Delta H \over n}.
\end{equation}
As a result, for a stoichiometric gas mixture the final HCl density
$n^\prime_{HCl}$ can be related to the residual H$_2$ density
$n^\prime_{H_2} = n^\prime_{Cl_2}$:
\begin{equation}
{n_{HCl}^{\prime\,2} \over n_{H_2}^\prime n_{Cl_2}^\prime} =
e^{\Delta H/k_B T}.
\end{equation}
This may be solved for the reaction (\ref{HClreact}) to yield
\begin{equation}
n^\prime_{HCl} = {n_{H_2} \over 1/2 + e^{-\Delta H/2 k_B T}},
\end{equation}
where an unprimed variable indicates the pre-reaction density.  After
solving for $T$ we find
\begin{equation}
n^\prime_{HCl} = 1.83 n_{H_2};
\end{equation}
about 8.6\% of the initial reactants remain unreacted.  This reduces the
energy released by a similar amount.

An analogous, but more complicated (because the number of molecules is
changed by the reaction) calculation can be performed for reaction
(\ref{HOreact}).  The fractional reduction in energy yield is greater than
for reaction (\ref{HClreact}) because the greater number of momentum states
accessible to the greater number of reactant molecules shifts the
equilibrium to the left.  This is analogous to the fact that a gas is 50\%
ionized at temperatures far below those corresponding to the ionization
energy.  Some results are reported in \cite{MMC60}, who found that for a
stoichiometric mixture H$_2$O constituted barely 50\% of the mole fraction
of the product gas.
\subsection{Frangible Gas Containment}
We typically want to apply an impulsive load to a test spherical shell, as
analyzed in Secs.~\ref{thinshell} and \ref{dynamic}.  This requires that
the outer shell containing the H$_2$--Cl$_2$ mixture rapidly release the
hot, high pressure, reaction products.  Their sound speed is about 1 km/s so
that an annulus 3 cm thick will decompress in tens of $\mu$s if the
containment shell is frangible and breaks into many small fragments.  This
is much less than the time over which the test object flows plastically.

Candidate materials are glass and the brittle plastics polystyrene and
PMMA, each of which have Weibull moduli $\sim$ 6--10.  Because they are
brittle it is probably necessary that their static, pre-explosion, loads
be several times less than their nominal tensile strengths $Y$, which are
about 500--800 bars.  For an internal pressure $P_i$ and an acceptable
stress $\sigma$ the required fractional thickness of a spherical shell
\begin{equation}
{\Delta R \over R} \ge {1 \over 2} {P_i \over \sigma}.
\end{equation}
Taking $P_i = 20$ bars, sufficient to produce a post-explosion pressure
$P_f \approx 200$ bars (Eq.~\ref{pHCl}) and a tolerable $\sigma = 200$
bars the required $\Delta R /R = 0.05$.  These parameters may be scaled
freely, but plausible values are $R = 5$ cm and $\Delta R = 2.5$ mm.  The
post-explosion pressure would produce, if confined, a static stress in the
shell of 2 kbar, far in excess of its strength and ensuring immediate
fragmentation.

In order to maintain symmetric pressure in the explosion products until
they have vented and their pressure has become insignificant, the shell
must fragment into pieces much smaller than the thickness of the annulus
between it and the test object.  This may be assured if the shell is scored
by grooves, perhaps 5 mm apart, to guide its fragmentation.  The grooves
should be rounded to avoid stress concentration and fracture before the
explosion.  Brittle fracture occurs when the stress intensity at a crack or
groove
\begin{equation}
K \approx \sigma \sqrt{\pi a},
\end{equation}
where $a$ is a crack depth, exceeds the fracture toughness $K_{IC}$.
For these plastics $K_{IC} \sim 10^8$ erg/cm$^{5/2}$.  It is necessary that
the material not fracture under the pressure of the reactant gas, so that
\begin{equation}
a < {K_{IC}^2 \over \pi \sigma^2} \lesssim 0.1\,\text{cm}.
\end{equation}
Because the properties of brittle materials vary from sample to sample, this
is only a rough guide, but establishes the feasibility of controlling the
fracture in such a manner as to insure breakup into many small fragments,
insuring approximately symmetric venting of the high pressure explosion
products.
\subsection{Numerical Example}
If the initial total pressure of a stoichiometric mix of reactants is
30 bars, then the pressure after explosion will be about 300 bars, and
the impulse delivered about 10 ktaps ($1 \times 10^4$ g/cm-s).  As discussed
in Sec.~\ref{plasticwork}, this is an interesting regime for inducing two
dimensional plastic flow in a thin sphere of ductile metal.  A more
quantitative result requires a code calculation, but this should be accurate
to a factor of two.
\section{Discussion}
Many different yield surfaces have been proposed, of which von Mises's is
probably the most popular for ductile materials.  Unfortunately, because of
the difficulty of measure compressive strength in any geometry other than
uniaxial compression, there are few empirical grounds for preferring one
yield surface to another.  This paper proposes a method of measuring
strength in two-dimensional compression that would provide an additional
empirical parameter and constrain the yield surfaces of the tested
materials.

The appendices discuss alternate means of applying impulsive loades to the
outside of a thin spherical shell.  These appear less promising, either
because of the difficulty of ensuring symmetric loading or of providing
sufficient impulse.  They are included for completeness.
\section{Acknowledgement}
This work was performed under the auspices of the U.S. Department of Energy
by Lawrence Livermore National Laboratory under Contract
DE-AC52-07NA27344 and by the Los Alamos National Laboratory, operated by
Triad National Security, LLC for the National Nuclear Security
Administration of the US Department of Energy under Contract
89233218CNA000001.
\section{Data Availability Statement}
This theoretical study produced no original data.
\appendix
\section{Light Initiated High Explosives (LIHE)}
\begin{figure}
\begin{center}
\includegraphics[width=5in]{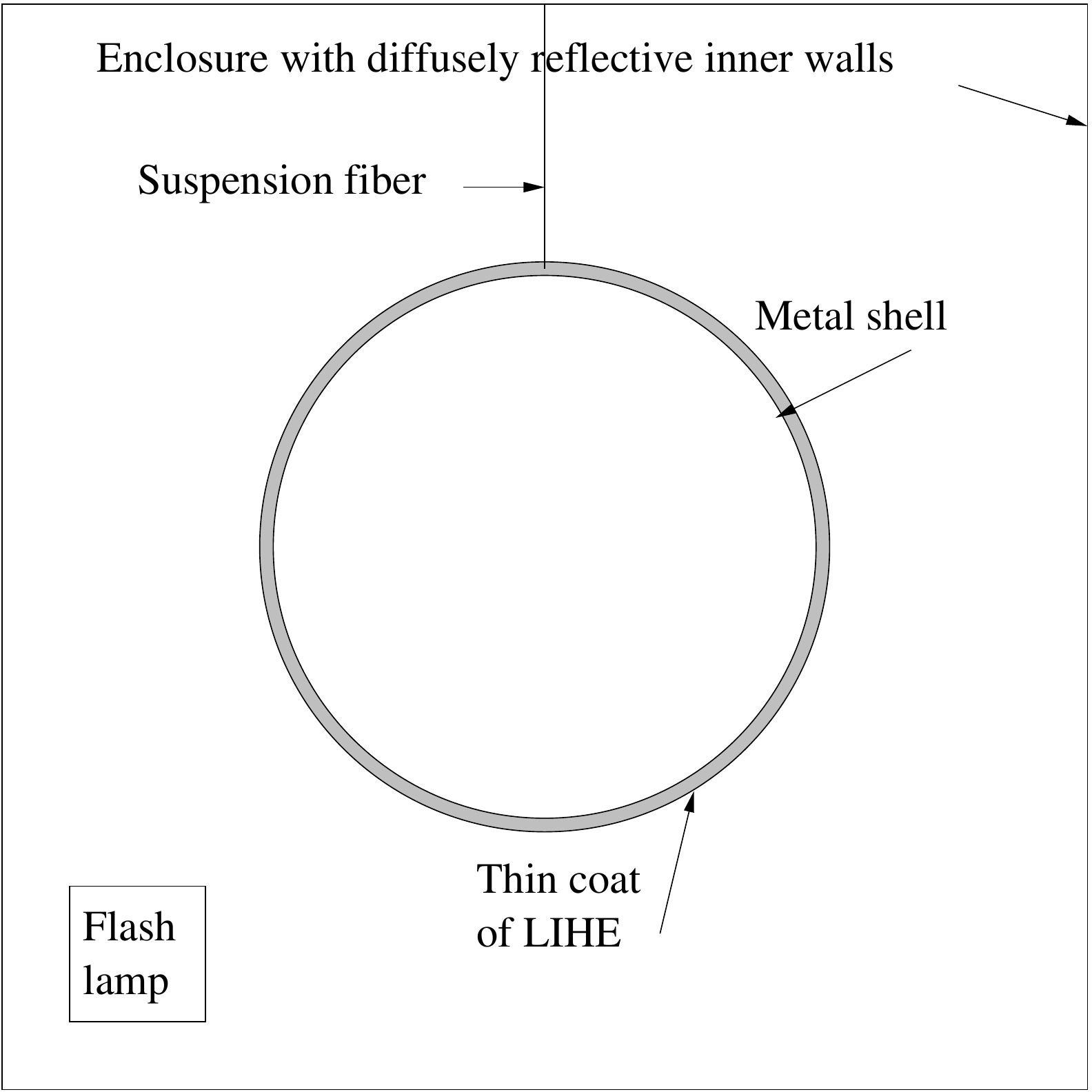}
\caption{Schematic of high strain rate LIHE biaxial compression experiment.}
\label{exptfig2}
\end{center}
\end{figure}

Light initiated high explosives \cite{B80a,B80b,FAS,C09}, deposited in thin
layers, are used to apply impulsive loads to the outside of a thin spherical
shell.  The thickness of the explosive layer is chosen so that it applies an
impulse per unit area $p_f$ to the shell.  The required impulses
(Eq.~\ref{pf}) are readily achievable. This is illustrated in
Fig.~\ref{exptfig2}.

The drawback of this method is that it is difficult to apply LIHE uniformly.
It is typically estimated that nonuniformities are $\sim 20\%$.  The effect
of the resulting asymmetry of the implosion would require two- (or three-)
dimensional elastic-plastic flow calculations, beyond the scope of this
paper, to evaluate.  In particular, if the flow were to develop hinges the
spherical description would become qualitatively invalid and results
difficult to interpret.  We have investigated the H$_2$-Cl$_2$ explosion
method chiefly because it may more readily apply a symmetric load to the
test object.  However, if the asymmetry problem can be overcome, the LIHE
method may be comparatively simple to implement.

\section{``Hydraulic'' Fluids}
Pressure might be applied with a ``hydraulic'' fluid of high bulk modulus and
sound speed, such as gallium or a gallium alloy of low melting 
point\footnote{Galistan, the ternary gallium-indium-tin eutectic with its
its eutectic point below room temperature, may be the best candidate.}.  The
fluid may be pressurized by the release of electrical or chemical energy or
by the impact of flyer plates. 

The fluid's high sound speed would relax spatially asymmetric transients if
they were strongly dissipative, but in the necessary regime of weak shocks
dissipation is negligible and these weak shocks would reverberate in the
``hydraulic'' fluid rather than relaxing to a uniform pressure, providing
asymmetric, rather than the spherically symmetric, compression required to
prevent dimpling or crumpling.  In addition, it is probably not possible to
make sufficiently small explosive charges (small critical radii imply
extreme sensitivity).
\section{Piezoelectric Drive}
Piezoelectric drive is relegated to this Appendix because saturation of
dielectric response limits the impulse that can be delivered to levels
insufficient to drive even a very soft metal like indium to plastic flow.
However, the analysis is of intrinsic interest.
\subsection{Symmetry}
Piezoelectric drive is spherically symmetric if the driving electric field
is produced between concentric spherical electrodes and if the piezoelectric
ceramic is radially and spherically symmetrically poled.  Spherically
symmetric poling may be achieved by poling hemispherical shells that are
later joined.  Because the relative dielectric constant of PZT ceramics is
$\sim 1000\text{--}2000$ the fringing fields at the hemispherical boundaries
during poling are very small, and the poling is nearly symmetric.
Alternatively, it might be possible to pole the entire spherical shell of
piezoelectric with a floating inner electrode if the electrical conductivity
of the piezoelectric is small enough.  Very small conductivities are hard
to measure, but extrapolation of conductivities measured at higher
temperature \cite{R10} to typical poling temperatures of 100--160 C suggests
that discharge times $t_{RC} = \epsilon/\sigma$, where here $\sigma$ is the
conductivity, may be as long as hours.  Here we outline the design of such a
piezoelectric-driven biaxial compression experiment.

\begin{figure}
\begin{center}
\includegraphics[width=5in]{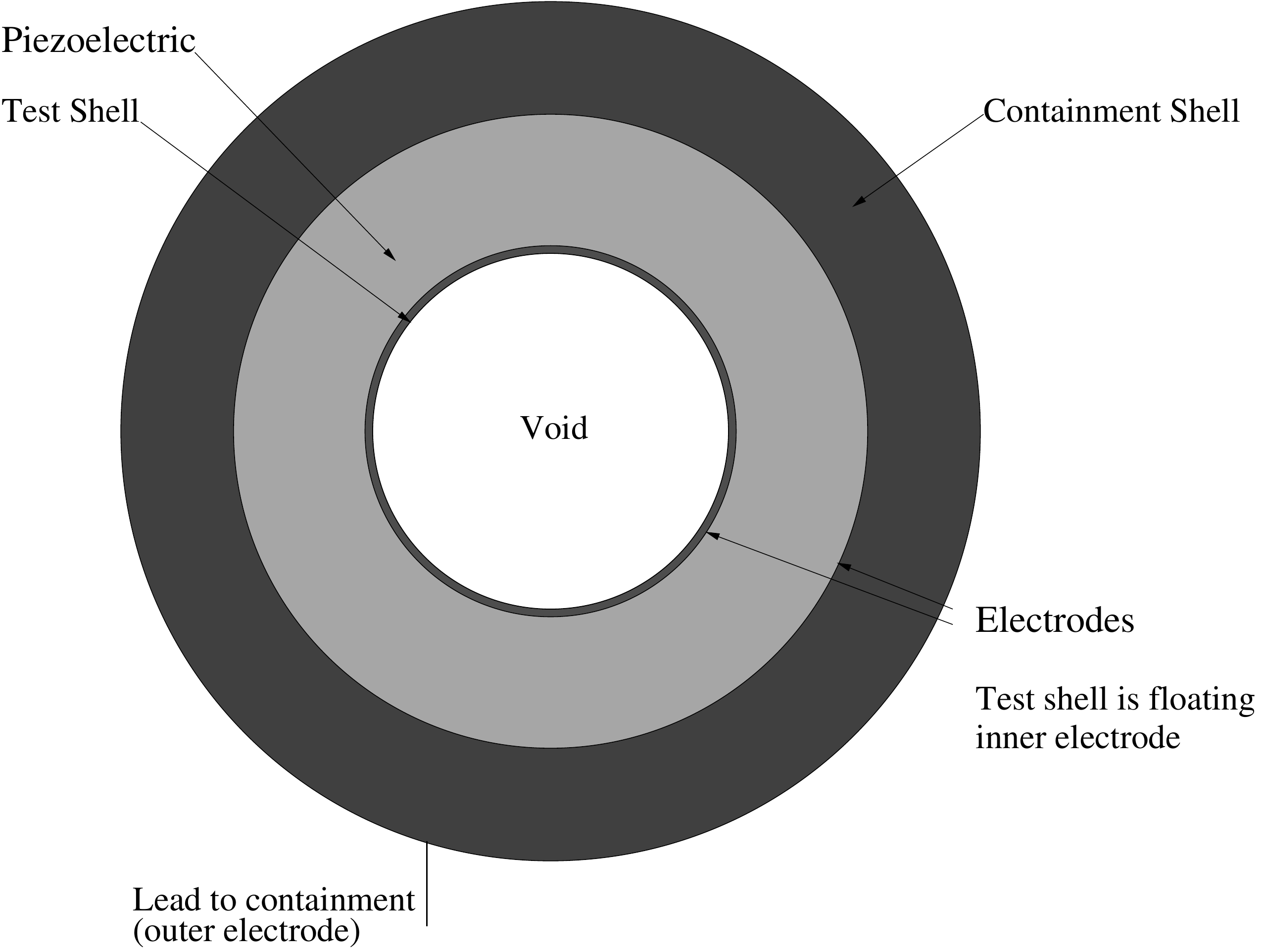}
\caption{Schematic of high strain rate piezoelectric compression apparatus.
The containment shell and piezoelectric are made in two pieces that may be
assembled before the shot,  and disassembled afterward (the piezoelectric
ceramic will likely shatter) to allow measurement of the plastic flow of the
test object.  The test shell and the containment shells are also the 
electrodes driving the piezoelectric.  Piezoelectrics like PZT are
sufficiently resistive that the test object/inner electrode can float
electrically over the time of the experiment; it is not necessary to
penetrate the piezoelectric with a lead wire to the test object/inner
electrode.}
\label{exptfig}
\end{center}
\end{figure}

Place a thin spherical shell of the test material, of density $\rho$, 
initial radius $r_0$ and thickness $\delta r$ inside a spherical, radially 
poled, piezoelectric shell.  Contain the piezoelectric within a stiff and
strong containment shell.  The containment shell serves several functions:
it provides one of the electrodes required to apply an electric field to the
piezoelectric, reflection of radial expansion of the piezoelectric by the
shell provides an additional source of compression of the test object, and
it contains the brittle piezoelectric if reflected pressure
pulses create a tensile stress sufficient to fragment it.  This is shown
in Figure~\ref{exptfig}.  

The electric discharge time of the floating inner electrode 
\begin{equation}
t_{discharge} = {\epsilon_0 \kappa \over \sigma},
\end{equation}
where $\kappa \approx 2000$ is the relative permittivity of the 
piezoelectric and $\sigma \approx 10^{-11}$ mho/m is its electrical 
conductivity \cite{R10}.  Then $t_{discharge} = {\cal O}(10^3)$ s, much
longer than the duration of the dynamic phase of the experiment, so that no
electric lead to the floating inner electrode is necessary.  This preserves
spherical symmetry.  Aside from simplifying calculation and analysis,
spherical symmetry reduces or eliminates any tendency of the test object to
dimple, hinge, or otherwise deviate from uniform flow, as it might if subject
to aspheric loading. 

\subsection{Equations}
The equations of a piezoelectric ceramic, in engineering notation in which
the subscript~3 denotes the radial (poled) direction, are
\begin{eqnarray}
S_1 &= &s_{11}T_1 + s_{12}T_2 + s_{13}T_3 + d_{13}E_3 \\
S_2 &= &s_{21}T_1 + s_{22}T_2 + s_{23}T_3 + d_{23}E_3 \\
S_3 &= &s_{31}T_1 + s_{32}T_2 + s_{33}T_3 + d_{33}E_3,
\end{eqnarray}
where $E_3$ is the radial electric field, the $S_i$ are the principal
components of strain, the $T_i$ are the principal components of stress, the
$s_{ij}$ are the components of the compliance tensor and the $d_{ij}$ are
piezoelectric coefficients.  For a ceramic in which the 1 and 2 directions
are equivalent $s_{11} = s_{22}$, $s_{12} = s_{21}$ and $s_{13} = s_{23}$.
The $s_{ij}$ are defined at constant electric field if the outer electrode
is maintained at constant potential with respect to ground during the pulse
that accelerates the test object (which is grounded by charge leakage prior
to the experiment).  The implied electric work done by or on the power
supply is not considered explicitly.

For an instantaneously applied electric field $S_i = 0$ because the material
has no time to move.  Solving for the stresses under these conditions, and
simplifying by substituting components of the elasticity tensor $c_{ij}$ for
convenience, we find
\begin{equation}
T_1 = T_2 = E_3 d_{13} \left[{1 \over s_{11} + s_{12}} + c_{13}
\left({d_{33} \over d_{13}} + 2{c_{13} \over c_{33}}\right)\right]
\end{equation}
and 
\begin{equation}
T_3 = - E_3 \left(d_{33}c_{33} + 2 c_{13}d_{13}\right).
\end{equation}
The stress drop $T_3 - 0$ between the inside and outside of the
test shell accelerates it inward.

The equation of motion of the test shell is
\begin{equation}
\label{pequation}
{dp \over dt} = T_3 - {c_{33} \over \delta r_{piezo}} \int\! {p \over \rho
\delta r}\,dt,
\end{equation}
where $p$ is its momentum per unit area and we have made the approximation
of only a small inward displacement, and therefore of constant thickness
$\delta r$ and radius $r_0$.  The second term arises from the reduction in
stress in the piezoelectric as the test shell moves inward, sending a
rarefaction into it, and makes the approximations that the piezoelectric
thickness $\delta r_{piezo} \ll r$ and that it responds quasistatically to
the motion of the test object.  The final approximation is justified by the
stiffness of ceramics: for PZT $c_{33} \approx 1.2 \times 10^{12}$
erg/cm$^3$ and the sound speed is several km/s.

Differentiating again with respect to time
\begin{equation}
{d^2p \over dt^2} = - b^2 p,
\end{equation}
where
\begin{equation}
b \equiv \sqrt{c_{33} \over \rho \delta r_{piezo} \delta r}.
\end{equation}
Using the boundary conditions $p = 0$ and ${dp \over dt} = T_{30}$ at
$t = 0$ we find the solution
\begin{equation}
\label{psolution}
p = {T_{30} \over b} \sin{bt}.
\end{equation}

Equation \ref{pequation} is only valid as long as the radial stress in the
piezoelectric remains compressive:
\begin{equation}
\Delta P = T_{30} - {c_{33} \over \delta r_{piezo}} \int\! {p \over
\rho \delta r} \ge 0.
\end{equation}
Once $T_3 < 0$ a void will open between the piezoelectric and the test
shell, and the effective $T_3 = 0$, to good approximation.  This happens
after a time $t_{void} = \pi /(2b)$.  After this time the test shell coasts
inward, until all its kinetic energy is dissipated in plastic work.

The final momentum per unit area
\begin{equation}
p_f = {T_{30} \over b}.
\end{equation}
Equating the kinetic energy per unit volume $p_f^2 /(2 \rho (\delta r)^2)$
to the plastic work (\ref{worksphere}) done before the shell stops after an
inward displacement $\Delta r \ll r_0$ yields a relation between $S_b$ and
$\Delta r$, with the other variables as parameters:
\begin{equation}
\label{yzero}
S_b = {r_0 \over \Delta r} {T_{30}^2 \delta r_{piezo} \over 4 \delta r
c_{33}}.
\end{equation}
Equivalently, we may write the required initial piezoelectric stress
\begin{equation}
\label{T30}
T_{30} = \sqrt{4 S_b c_{33}{\Delta r \over r_0}{\delta r \over \delta
r_{piezo}}}.
\end{equation}

Inverting Eq.~\ref{yzero} yields estimates of useful ranges of the 
parameters.  For a nominal transition metal (like copper or tin) of density
$\rho = 9$ gm/cm$^3$ with $\delta r = 0.1$ cm and $\delta r_{piezo} = 0.3$
cm, the rate parameter $b = 2.1 \times 10^6$/s.  The strain rate varies
during the trajectory of the shell, but may be approximated
\begin{equation}
{\dot \epsilon} \approx {\epsilon \over t_{void}} \approx {4 \over \pi}
{\Delta r \over r_0} b \approx 2.7 \times 10^5/{\rm s},
\end{equation}
where the radial strain $\epsilon = 2 \Delta r/r_0 = 0.2$; the strains and
strain rates in the two tangential dimensions are half as large.

The value of $T_{30}$ required to achieve this $\Delta r$, strain and
strain rate depends on the strength $S_b$; conversely, the strength inferred
from an experimental measurement of $\Delta r$ depends on $T_{30}$.  As a
numerical example, if $S_b = 7 \times 10^8$ erg/cm$^3$ (700 bars), a typical
value for pure copper, then for the preceding parameters $T_{30} = 1.06
\times 10^{10}$ erg/cm$^3$ (10.6 kbar).  In practice, smaller values of 
$T_{30}$ are required for smaller $\Delta r /r_0$ and $\delta r/\delta 
r_{piezo}$; these are experimentally feasible, but if $\delta r_{piezo}
\centernot\ll r_0$ then the simple analytic result is inapplicable.  It should
also be remembered that $S_b$ will be affected by work hardening if $\delta
r/r_0$ is not very small; in practice, strength models incorporating work
hardening and thermal effects and a quantitative treatment of the elastic
and piezoelectric response would be required and numerically integrated.
\subsection{Power Supply}
The piezoelectric equations yield the relation between the radial electric
field $E_3$ and the normal component of stress $T_3$ on a spherical surface:
\begin{equation}
\label{E3}
E_3 = - {T_3 \over d_{33} c_{33} + 2 d_{13} c_{13}}.
\end{equation}
Substituting typical values of the elastic and piezoelectric coefficients
for PZT yields the estimate (in MKS units, for convenience)
\begin{equation}
E_3 \approx - T_3 \times 0.08 {\rm V \over \text{m-Pa}}.
\end{equation}
To achieve a nominal $T_3 = 10^7$ Pa (100 bars) requires $E_3 \approx 10^6$
V/m, a value comparable to coercive field of PZT piezoelectrics, and
approximately the maximum field at which their piezoelectric response
remains unsaturated.  At high frequencies the coercive field may be higher,
but is likely to be of the same order of magnitude.  The potential drop
across the previously assumed shell of piezoelectric 0.3 cm thick is then
$V = 3$ kV.

The capacitance of the piezoelectric is
\begin{equation}
C = 4 \pi \kappa \epsilon_0 {r_0 (r_0 + \delta r_{piezo}) \over \delta
r_{piezo}} \approx 0.5\,\mu{\rm F},
\end{equation}
where we have taken the relative dielectric constant (at zero strain) of PZT
$\kappa_{33} = 1000$, a representative value.  The energy required to charge
it to the required potential
\begin{equation}
{\cal E} = {C V^2 \over 2} \approx 2.2\,{\rm J}.
\end{equation}

Charging must be accomplished in a time $t_{charge} \ll 1/b$, and implies a
current 
\begin{equation}
J \simeq {CV \over t_{charge}} \gg CVb \simeq 3 \times 10^3\,{\rm Amp},
\end{equation}
or a charging power
\begin{equation}
P \simeq JV \gg {CV^2 b \over 2} \simeq 5 \times 10^6\,{\rm W}.
\end{equation}
The implied impedance
\begin{equation}
Z \equiv {V \over J} \ll {1 \over C b} \simeq 1\,\Omega.
\end{equation}
These parameters are not challenging for a high voltage capacitive
pulsed power supply.
\subsection{Transient Piezoelectric Response}
Piezoelectrics are described by their static displacement coefficients
$d_{33}$ describing the ratio of strain parallel to electric field applied
in the poling direction (denoted by 3), $d_{31} = d_{32}$ (negative) the
ratio of strain in the transverse (1 and 2) directions to the electric
field applied in the poling direction, and $d_{15}$ the ratio of shear in
the 13 plane to electric field applied in the transverse direction.  Symmetry
implies values for other coefficients.  This is a quasi-static description.

We are concerned with fields applied much faster than the elastic relaxation
time of the piezoelectric.  When such a field is applied, there is an
immediate increase in stress, but strain follows only more slowly, on the
time scale of the acoustic transit time across the piezoelectric, and in
general the deformation is oscillatory as the initial stress distribution
excites elastic waves.  The immediate stress is the stress field that would
return the quasi-statically strained configuration to its unstrained (zero
electric field) state.

The quasi-static strain is
\begin{equation}
\mbox{\boldmath $u$} =
\begin{vmatrix}
d_{13}E_3 & 0 & 0 \\
0 & d_{13}E_3 & 0 \\
0 & 0 & d_{33}E_3
\end{vmatrix},
\end{equation}
where the electric field ${\vec E} = E_3 {\hat z}$.  Using the relation
between stress and strain,
\begin{equation}
\mbox{\boldmath $\sigma$}_{ik} = K u_{ll} \delta_{ik} + 2 \mu \left(u_{ik}
- {1 \over 3} \delta_{ik} u_{ll}\right),
\end{equation}
where $K$ is the bulk modulus and $\mu$ the shear modulus.

We first consider an unconstrained piezoelectric.  Then the stress state
that produces these strains is
\begin{eqnarray}
\sigma_{22} = \sigma_{11} &= K(2d_{13}+d_{33})E_3 + 
2\mu \left({1 \over 3}d_{13} + {1 \over 3}d_{33}\right)E_3\\
\sigma_{33} &= K(2d_{13}+d_{33})E_3 +
2\mu \left({2 \over 3}d_{33} - {2 \over 3}d_{13}\right)E_3,
\end{eqnarray}
where $E_3$ is the magnitude of the electric field in the 3-direction.

If the constraint $u_{11} = u_{22} = 0$ is imposed, as appropriate for a
spherical shell in which the 3-direction of poling and electrification is
radial, an additional stress field $\mathbf \sigma^\prime$ must be added:
\begin{eqnarray}
\sigma_{22}^\prime = \sigma_{11}^\prime &= d_{13}E_3
\left[2K(\nu-1)+2\mu\left(-{1 \over 3}-{2 \over 3}\nu\right)\right]\\
\sigma_{33}^\prime &= d_{13}E_3
\left[2K(\nu-1)+2\mu\left({2 \over 3}+{4 \over 3}\nu\right)\right].
\end{eqnarray}

Combining these, the radial component of stress that acts on the shell
pushed by the piezoelectric
\begin{equation}
\sigma_{33} = KE_3(2 \nu d_{13}+d_{33}) +
2\mu E_3\left({4 \over 3}\nu d_{13}+{2 \over 3}d_{33}\right).
\end{equation}
\subsection{Estimates}
Using Eqs.~\ref{T30} and \ref{E3} we estimate the electric field, assuming
linear piezoelectric response, required to produce significant plastic flow.
We take an inward plastic displacement of the test sample $\Delta r = 0.3
r_0$ and a sample thickness $\delta r = 0.3 \delta r_{piezo}$; for example,
a sample with initial radius $r_0 = 1$ cm and initial thickness $\delta r =
1$ mm surrounded by a piezoelectric $r_{piezo} = 3$ mm thick.  For
representative parameters of PZT piezoelectric ceramic and a test sample of
copper with $S_b = 7 \times 10^8$ erg/cm$^3$ (700 bars) we require $E_3 = 3
\times 10^7$ V/m, while even for very soft indium with $S_b = 9 \times 10^6$
erg/cm$^3$ (9 bars) \cite{R88} we require $E_3 = 3.6 \times 10^6$ V/m.  Both
these values far exceed the fields of about $5 \times 10^5$ V/m at which
piezoelectrics saturate and depole.  Piezoelectric drive cannot be strong
enough to plastically deform even very soft metals.

\end{document}